\documentclass[pra,twocolumn,floatfix,aps,superscriptaddress,showpacs,amsfonts]{revtex4}
\usepackage{graphicx}
\usepackage{array}
\usepackage{amsthm}
\usepackage{amsmath}
\usepackage{amssymb}

\begin{document}

\title{Monogamy of $\alpha$th Power Entanglement Measurement in Qubit Systems}
\author{Yu Luo}
\author{Yongming Li}
\affiliation{College of Computer Science, Shaanxi Normal University, Xi'an, 710062, China}
\date{\today}

\begin{abstract}
In this paper, we study the $\alpha$th power monogamy properties related to the entanglement measure in bipartite states. The monogamy relations related to the $\alpha$th power of negativity and the Convex-Roof Extended Negativity are obtained for $N$-qubit states. We also give a tighter bound of hierarchical monogamy inequality for the entanglement of formation. We find that the GHZ state and W state can be used to distinguish both the $\alpha$th power of the concurrence for $0<\alpha<2$ and the $\alpha$th power of the entanglement of formation for $0<\alpha\leq\frac{1}{2}$. Furthermore, we compare concurrence with negativity in terms of monogamy property and investigate the difference between them.
\end{abstract}
\eid{identifier}
\pacs{03.67.a, 03.65.Ud, 03.65.Ta}
\maketitle

\section{Introduction}
 Multipartite entanglement is an important physical resource in quantum mechanics, which can be used in quantum computation, quantum communication and quantum cryptography. One of the most surprising phenomenon for multipartite entanglement is the monogamy property, which may be as fundamental as the no-cloning theorem~\cite{Bruss99,Coffman00,Osborne06,Kay09}. The monogamy property can be interpreted as the amount of entanglement between $A$ and $B$, plus the amount of entanglement between $A$ and $C$, cannot be greater than the amount of entanglement between $A$ and the pair $B$$C$. Monogamy property have been considered in many areas of physics: one can estimate the quantity of information captured by an eavesdropper about the secret key to be extracted in quantum cryptography~\cite{Osborne06,Barrett05}, the frustration effects observed in condensed matter physics~\cite{Ma11}, even in black-hole physics~\cite{Susskind13,Lloyd14}.

Historically, monogamy property of various entanglement measure have been discovered. Coffman $et$ $al$ first considered three qubits $A$,$B$ and $C$ which may be entangled with each other~\cite{Coffman00}, who showed that the squared concurrence $\mathcal{C}^2$ follows this monogamy inequality. Osborne $et$ $al$ proved the squared concurrence follows a general monogamy inequality for $N$-qubit system~\cite{Osborne06}. Analogous to the Coffman-Kundu-Wootters (CKW) inequality, Ou $et$ $al$ proposed the monogamy inequality holds in terms of squared negativity $\mathcal{N}^2$~\cite{Ou07}. Kim $et$ $al$ showed that the squared convex-roof extended negativity $\mathcal{\widetilde{N}}^2$ follows monogamy inequality~\cite{Kim09}. Oliveira $et$ $al$ and Bai $et$ $al$ investigated entanglement of formation(EoF) and showed that the squared EoF $E^2$ follows the monogamy inequality ~\cite{Oliveira14,Bai1401}. A natural question is why those monogamy property above are squared entanglement measure? In fact, Zhu $et$ $al$ showed that the $\alpha$th power of concurrence $\mathcal{C}^{\alpha}$ ($\alpha\geq2$) and the $\alpha$th power of entanglement of formation $E^{\alpha}$ ($\alpha\geq\sqrt2$) follow the general monogamy inequalities~\cite{Zhu14}. Sometimes, we can view the coefficient $\alpha$ as a kind a of assigned weight to regulate the monogamy property~\cite{Regula14,Salini14}.

In this paper, we study the monogamy relations related to $\alpha$th power of some entanglement measures. We show that the $\alpha$th power of negativity $\mathcal{N}^{\alpha}$ and the $\alpha$th power of convex-roof extended negativity (CREN) $\mathcal{\widetilde{N}}^{\alpha}$ follows the hierarchical monogamy inequality for $\alpha\geq2$~\cite{Bai1402}. From the hierarchical monogamy inequality, the general monogamy inequalities related to $\mathcal{N}^{\alpha}$ and $\mathcal{\widetilde{N}}^{\alpha}$ are obtained for $N$-qubit states. We find that the GHZ state and W state can be used to distinguish the $\mathcal{C}^{\alpha}$ for $0<\alpha<2$, which situation was not clear in Zhu $et$ $al$'s paper~\cite{Zhu14}. We also prove the GHZ state and W state can be used to distinguish both the $\alpha$th power of EoF for $0<\alpha\leq\frac{1}{2}$. The hierarchical monogamy inequality for $E^{\alpha}$ is also discussed, which improved Bai $et$ $al$'s result~\cite{Bai1402,Bai1401}.

This paper is organized as follows. In Sec.~\ref{sec:Neg},we study the monogamy property of $\alpha$th power of negativity. In Sec.~\ref{sec:CREN}, we discuss the monogamy property of $\alpha$th power of CREN. In Sec.~\ref{sec:EoF}, we study the monogamy property of $\alpha$th power of EoF. In Sec.~\ref{sec:Vs}, we compare the monogamy property of concurrence with negativity. We summarize our results in Sec.~\ref{sec:conclusion}.

\section{Monogamy of $\alpha$th power of Negativity}\label{sec:Neg}
Given a bipartite state $\rho_{AB}$ in the Hilbert space $\mathcal{H_A}\otimes\mathcal{H_B}$. Negativity is defined as~\cite{Vidal02}:
\begin{equation}
\mathcal{N}(\rho_{AB})=\frac{\|\rho^{T_A}_{AB}\|-1}{2},\label{eq:Negativity}
\end{equation}
where $\rho^{T_A}_{AB}$ is the partial transpose with respect to the subsystem $A$, $\|X\|$denotes the trace norm of $X$, i.e $\|X\|\equiv Tr\sqrt{XX^{\dag}}.$ Negativity is a $computable$ measure of entanglement, and which is a convex function of $\rho_{AB}$. $\mathcal{N}(\rho_{AB})=0$ if and only if $\rho_{AB}$ is separable for the $2\otimes2$ and $2\otimes3$ systems~\cite{Horodecki98}. For the purposes of discussion , we use following definition of negativity:
\begin{equation}
\mathcal{N}(\rho_{AB})=\|\rho^{T_A}_{AB}\|-1.\label{eq:Negativity}
\end{equation}
For any maximally entangled state in two-qubit system, this definition of negativity is equal to 1.

For a bipartite pure state $|\psi_{AB}\rangle$, the concurrence is defined as:
\begin{equation}\label{eq:Conpure}
\mathcal{C}(|\psi_{AB}\rangle)=\sqrt{2[1-Tr(\rho_A^2)]}=2\sqrt{\det\rho_A},
\end{equation}
where $\rho_A$ is the reduced density matrix of subsystem A. For a mixed state $\rho_{AB}$, the concurrence can be defined as:
\begin{equation}\label{eq:Conmixed}
\mathcal{C}(\rho_{AB})=\min\sum_ip_i\mathcal{C}(|\psi_{AB}^i\rangle),
\end{equation}
where the minimum is taken over all possible pure state decompositions $\{p_i,\psi_{AB}^i\}$ of $\rho_{AB}.$

The next $lemma$ builds a relationship between negativity and concurrence in a $2\otimes m\otimes n$ system ($m\geq2,n\geq2$).

{\sf Lemma 1}~. For a pure state $|\psi\rangle_{ABC}$ in a $2\otimes m\otimes n$ system ($m\geq2,n\geq2$), the negativity of bipartition $A|BC$ is equal to its concurrence: $\mathcal{N}_{A|BC}=\mathcal{C}_{A|BC}$, where $\mathcal{N}_{A|BC}=\mathcal{N}(|\psi_{ABC}\rangle)$ and $\mathcal{C}_{A|BC}=\mathcal{C}(|\psi_{ABC}\rangle).$

\emph{Proof:} Based on the Schmidt decomposition, we can write the bipartition $A|BC$ as: $|\psi_{A|BC}\rangle=\sum_i\sqrt{\lambda_{i}}|\phi^{i}_A\rangle\otimes|\phi^{i}_{BC}\rangle$, where $\lambda_i$ are Schmidt coefficients and $\sum_i\lambda_i=1$. $\{|\phi^{i}_A\rangle\}$,$\{|\phi^{i}_{BC}\rangle\}$ are orthogonal basis for system $A$ and system $BC$ respectively. The density operator $\rho_{ABC}=\sum_{i,j}\sqrt{\lambda_{i}\lambda_{j}}|\phi^{i}_A\rangle\langle\phi^{j}_A|\otimes|\phi^{i}_{BC}\rangle\langle\phi^{j}_{BC}|$, the partial transpose of $\rho_{ABC}$ with respect to system A is given by: $\rho^{T_A}_{ABC}=\sum_{i,j}\sqrt{\lambda_{i}\lambda_{j}}|\phi^{j\ast}_A\rangle\langle\phi^{i\ast}_A|
\otimes|\phi^{i}_{BC}\rangle\langle\phi^{j}_{BC}|$. The negativity of $|\psi\rangle_{A|BC}$ is:
\begin{eqnarray}
\mathcal{N}_{A|BC}&=&\|\rho^{T_A}_{ABC}\|-1\nonumber\\
&=&\|\sum_{i,j}\sqrt{\lambda_{i}\lambda_{j}}|\phi^{j\ast}_A\rangle\langle\phi^{i\ast}_A|\otimes|\phi^{i}_{BC}\rangle\langle\phi^{j}_{BC}|\|-1\nonumber\\
&=&\|\sum_{i,j}\sqrt{\lambda_{i}\lambda_{j}}|\phi^{j\ast}_A\rangle\langle\phi^{j}_{BC}|\otimes|\phi^{i}_{BC}\rangle\langle\phi^{i\ast}_A|\|-1\nonumber\\
&=&\|\sum_{j}\sqrt{\lambda_{j}}|\phi^{j\ast}_A\rangle\langle\phi^{j}_{BC}|\otimes\sum_{i}\sqrt{\lambda_{i}}|\phi^{i}_{BC}\rangle\langle\phi^{i\ast}_A|\|-1\nonumber\\
&=&\|R\otimes R^{\dag}\|-1\nonumber\\
&=&\|R\|^2-1\nonumber\\
&=&(\sum_i\sqrt{\lambda_i})^2-1\nonumber\\
&=&2\sqrt{\lambda_0\lambda_1}\nonumber\\
&=&2\sqrt{\det\rho_A}\nonumber\\
&=&\mathcal{C}_{A|BC},
\end{eqnarray}
where $R=\sum_{j}\sqrt{\lambda_{j}}|\phi^{j\ast}_A\rangle\langle\phi^{j}_{BC}|$, and we have used the property of trace norm: $\|A\otimes B\|=\|A\|\otimes\|B\|.$ \qquad $\square$

Now we will study the monogamy property of $\alpha$th power of negativity $\mathcal{N}^{\alpha}$.

{\sf Theorem 1}~. For a pure state $|\psi_{A|B\textbf{C}}\rangle$ in a $2\otimes2\otimes2^{N-2}$ system,
the $\alpha$th power of negativity satisfies the monogamy inequality:
\begin{equation}
\mathcal{N}^{\alpha}_{A|B\textbf{C}}\geq\mathcal{N}^{\alpha}_{AB}+\mathcal{N}^{\alpha}_{A\textbf{C}},
\end{equation}
for $\alpha\geq 2,$
and satisfy the polygamy inequality:
\begin{equation}
\mathcal{N}^{\alpha}_{A|B\textbf{C}}<\mathcal{N}^{\alpha}_{AB}+\mathcal{N}^{\alpha}_{A\textbf{C}},
\end{equation}
for $\alpha\leq 0.$

\emph{Proof:} When $\alpha\geq2$, by using $Lemma$ $1$, we obtain $\mathcal{N}_{A|B\textbf{C}}=\mathcal{C}_{A|B\textbf{C}}.$ Combine with the result from Re.~\cite{Zhu14}:
\begin{equation}\label{eq:Concurrence}
\mathcal{C}^{\alpha}_{A|B\textbf{C}}\geq \mathcal{C}^{\alpha}_{AB}+\mathcal{C}^{\alpha}_{A\textbf{C}},
\end{equation}
for $\alpha\geq2$, We have
\begin{eqnarray}
\mathcal{N}^{\alpha}_{A|B\textbf{C}}&=&\nonumber\mathcal{C}^{\alpha}_{A|B\textbf{C}}\\\nonumber
&\geq& \mathcal{C}^{\alpha}_{AB}+\mathcal{C}^{\alpha}_{A\textbf{C}}\\
&\geq& \mathcal{N}^{\alpha}_{AB}+\mathcal{N}^{\alpha}_{A\textbf{C}},
\end{eqnarray}
 the last inequality is due to for any mixed state in a $2\otimes d$ $(2\leq d)$ quantum system, concurrence is an upper bound of negative, i.e. $\mathcal{N}_{A\textbf{C}}\leq \mathcal{C}_{A\textbf{C}}$~\cite{Chen05}. When $\alpha\leq0,$ without loss of generality, assuming $\mathcal{N}_{AB}\geq\mathcal{N}_{A\textbf{C}}>0,$ we have:
$\mathcal{N}^{\alpha}_{A|B\textbf{C}}\leq (\mathcal{N}^{2}_{AB}+\mathcal{N}^{2}_{A\textbf{C}})^{\frac{\alpha}{2}}=\mathcal{N}^{\alpha}_{AB}
(1+\frac{\mathcal{N}^{2}_{A\textbf{C}}}{\mathcal{N}^{2}_{AB}})^{\frac{\alpha}{2}}<\mathcal{N}^{\alpha}_{AB}
[1+(\frac{\mathcal{N}^{2}_{A\textbf{C}}}{\mathcal{N}^{2}_{AB}})^{\frac{\alpha}{2}}]=\mathcal{N}^{\alpha}_{AB}+\mathcal{N}^{\alpha}_{A\textbf{C}},$ where we used the property for the second inequality: $(1+x)^t<1+x^t  (x>0,t\leq0).$ If $\mathcal{N}_{AB}=0$ or $\mathcal{N}_{A\textbf{C}}=0,$ the inequality $\mathcal{N}^{\alpha}_{A|B\textbf{C}}<\mathcal{N}^{\alpha}_{AB}+\mathcal{N}^{\alpha}_{A\textbf{C}}$ obviously holds.
\qquad \qquad \qquad \qquad \qquad \qquad \qquad $\square$

If we consider any $N$-qubit pure state $|\psi_{A_1A_2\ldots{A_N}}\rangle$ in $k$-partite cases with $k=\{3,4,\ldots,N\}.$
From $Theorem$ $1$, a set of hierarchical monogamy inequalities of $\mathcal{N}^{\alpha}$ holds:
\begin{equation}
\mathcal{N}^{\alpha}_{A_1|A_2\ldots{A_N}}\geq\sum_{i=2}^{k-1}\mathcal{N}^{\alpha}_{A_1A_i}+\mathcal{N}^{\alpha}_{A_1|A_k\ldots{A_N}},
\end{equation}
for $\alpha\geq 2,$ and a set of hierarchical polygamy inequalities of $\mathcal{N}^{\alpha}$ holds:
\begin{equation}
\mathcal{N}^{\alpha}_{A_1|A_2\ldots{A_N}}\leq\sum_{i=2}^{k-1}\mathcal{N}^{\alpha}_{A_1A_i}+\mathcal{N}^{\alpha}_{A_1|A_k\ldots{A_N}},
\end{equation}
for $\alpha\leq 0.$

These set of hierarchical relations can be used to detect the multipartite entanglement in these $k$-partite~\cite{Bai1402}. We can also obtain the following result:

{\sf Corollary 1}~. For any $N$-qubit pure state $|\psi_{A_1A_2\ldots{A_N}}\rangle$ the general monogamous inequality hold:
\begin{equation}\label{eq:Corollary 1}
\mathcal{N}^{\alpha}_{A_1|A_2\cdots{A_N}}\geq \mathcal{N}^{\alpha}_{A_1A_2}+...+\mathcal{N}^{\alpha}_{A_1A_{N}},
\end{equation}
for $\alpha\geq 2,$ and the general polygamous inequality holds:
\begin{equation}
\mathcal{N}^{\alpha}_{A_1|A_2\cdots{A_N}}< \mathcal{N}^{\alpha}_{A_1A_2}+...+\mathcal{N}^{\alpha}_{A_1A_{N}},
\end{equation}
for $\alpha\leq 0.$

We can see the result of $\alpha=2$ from Re.~\cite{Ou07} is a special case of our monogamy inequality Eq.~(\ref{eq:Corollary 1}).

\section{Monogamy of $\alpha$th power Convex-Roof Extended Negativity}\label{sec:CREN}
 Given a bipartite state $\rho_{AB}$ in the Hilbert space $\mathcal{H_A}\otimes\mathcal{H_B}$. CREN is defined as the convex roof extended of negativity on pure states~\cite{Lee03}:

\begin{equation}
\mathcal{\widetilde{N}}(\rho_{AB})=\min\sum_ip_i\mathcal{N}(|\psi_{AB}^i\rangle),
\end{equation}
where the minimum is taken over all possible pure state decompositions $\{p_i,\psi_{AB}^i\}$ of $\rho_{AB}.$
Obviously, the CREN of a pure state is equal to its Negativity.  CREN gives a perfect discrimination of PPT bound entangled states and separable states in any bipartite quantum systems~\cite{Horodeki97,Dur00}. We have following result for CREN:

{\sf Theorem 2}~. For a mixed state $\rho_{A|B\textbf{C}}$ in a $2\otimes2\otimes2^{N-2}$ system,
the following monogamy inequality holds:
\begin{equation}
\mathcal{\widetilde{N}}^{\alpha}_{A|B\textbf{C}}\geq\mathcal{\widetilde{N}}^{\alpha}_{AB}+\mathcal{\widetilde{N}}^{\alpha}_{A\textbf{C}},
\end{equation}
for $\alpha\geq 2,$
and following polygamy inequality holds:
\begin{equation}
\mathcal{\widetilde{N}}^{\alpha}_{A|B\textbf{C}}<\mathcal{\widetilde{N}}^{\alpha}_{AB}+\mathcal{\widetilde{N}}^{\alpha}_{A\textbf{C}},
\end{equation}
for $\alpha\leq 0.$

\emph{Proof:} We only prove the first monogamy inequality, the proof of second inequality is similar to the proof of $Theorem$ $1$. Assuming a mixed state $\rho_{A|B\textbf{C}}$ in a $2\otimes2\otimes2^{N-2}$ system, by using the $Lemma$ $1$, the definition of CREN and concurrence, we have:
\begin{eqnarray}\label{CREN}
\mathcal{\widetilde{N}}_{A|B\textbf{C}}&=&\nonumber\min\sum_ip_i\mathcal{N}(|\psi_{A|B\textbf{C}}^i\rangle)\\\nonumber
&=&\min\sum_ip_i\mathcal{C}(|\psi_{A|B\textbf{C}}^i\rangle)\\
&=&\mathcal{C}_{A|B\textbf{C}}.
\end{eqnarray}
Thus we have:
\begin{eqnarray}
\mathcal{\widetilde{N}}^{\alpha}_{A|B\textbf{C}}&=&\nonumber\mathcal{C}^{\alpha}_{A|B\textbf{C}}\\\nonumber
&\geq&\mathcal{C}^{\alpha}_{AB}+\mathcal{C}^{\alpha}_{A\textbf{C}}\\
&\geq&\mathcal{\widetilde{N}}^{\alpha}_{AB}+\mathcal{\widetilde{N}}^{\alpha}_{A\textbf{C}},
\end{eqnarray}
for $\alpha\geq2$,
where the second inequality is due to for any mixed state in a $2\otimes d$ $(2\leq d)$ quantum system, concurrence is an upper bound of negative. \qquad \qquad \qquad \qquad $\square$

From $Theorem$ $2$, a set of hierarchical monogamy inequalities of $\mathcal{\widetilde{N}}^{\alpha}$ holds for any $N$-qubit mixed state $\rho_{A_1A_2\cdot{A_N}}$ in $k$-partite cases with $k=\{3,4,\ldots,N\}$:
\begin{equation}
\mathcal{\widetilde{N}}^{\alpha}_{A_1|A_2\ldots{A_N}}\geq\sum_{i=2}^{k-1}\mathcal{\widetilde{N}}^{\alpha}_{A_1A_i}+\mathcal{\widetilde{N}}^{\alpha}_{A_1|A_k\ldots{A_N}},
\end{equation}
for $\alpha\geq 2,$ and a set of hierarchical polygamy inequalities of $\mathcal{\widetilde{N}}^{\alpha}$ holds:
\begin{equation}
\mathcal{\widetilde{N}}^{\alpha}_{A_1|A_2\ldots{A_N}}\leq\sum_{i=2}^{k-1}\mathcal{\widetilde{N}}^{\alpha}_{A_1A_i}+\mathcal{\widetilde{N}}^{\alpha}_{A_1|A_k\ldots{A_N}},
\end{equation}
for $\alpha\leq 0.$

We also have the following corollary:

{\sf Corollary 2}~. For a mixed state $\rho_{A_1A_2\ldots A_N}$ in a $N$-qubit system, the $\alpha$th power of CREN satisfies:
\begin{equation}\label{eq:Corollary 2}
\mathcal{\widetilde{N}}^{\alpha}_{A_1|A_2\ldots A_N}\geq \mathcal{\widetilde{N}}^{\alpha}_{A_1A_2}+...+\mathcal{\widetilde{N}}^{\alpha}_{A_1A_{N}},
\end{equation}
for $\alpha\geq2$
and
\begin{equation}
\mathcal{\widetilde{N}}^{\alpha}_{A_1|A_2\ldots A_N}< \mathcal{\widetilde{N}}^{\alpha}_{A_1A_2}+...+\mathcal{\widetilde{N}}^{\alpha}_{A_1A_{N}}, \end{equation}
for $\alpha\leq0$.

We can see the result of $\alpha=2$ from Re.~\cite{Kim09} is a special case of our monogamy inequality Eq.~(\ref{eq:Corollary 2}).

\section{Monogamy of $\alpha$th power of Entanglement of Formation}\label{sec:EoF}
Given a bipartite state $\rho_{AB}$ in the Hilbert space $\mathcal{H_A}\otimes\mathcal{H_B}$, the entanglement of formation (EoF) is defined as~\cite{Bennett9601,Bennett9602}:
\begin{equation}
E(\rho_{AB})=\min\sum_ip_iE(|\psi^i_{AB}\rangle),
\end{equation}
where $E(|\psi^i_{AB}\rangle)=-Tr\rho_A^i\log_2\rho_A^i=-Tr\rho_B^i\log_2\rho_B^i$ is the von Neumann entropy, the minimum is taken over all possible pure state decompositions $\{p_i,\psi_{AB}^i\}$ of $\rho_{AB}.$ In Re.~\cite{Wootters98}, Wootters derived an analytical formula for a two-qubit mixed state $\rho_{AB}$:
\begin{equation}
E(\rho_{AB})=h(\frac{1+\sqrt{1-\mathcal{C}_{AB}^2}}{2}),
\end{equation}
where $h(x)=-x\log_2x-(1-x)\log_2(1-x)$ is the binary entropy and $\mathcal{C}_{AB}$ is the concurrence of $\rho_{AB}$ which is given by Eq.~(\ref{eq:Conpure}) and Eq.~(\ref{eq:Conmixed}).
Bai $et$ $al$ have proven a set of hierarchical monogamy inequalities holds for the squared EoF in a $2\otimes2\otimes2^{N-2}$ system~\cite{Bai1402}.
\begin{equation}
E^2_{A_1|A_2\ldots{A_N}}\geq\sum_{i=2}^{k-1}E^2_{A_1A_i}+E^2_{A_1|A_k\ldots{A_N}}.
\end{equation}
We will show that the hierarchical monogamy inequality holds for the $\alpha$th power of EoF, where $\alpha\geq\sqrt 2.$ Our result can be seen an improvement of Bai $et$ $al$'s work.

{\sf Theorem 3}~. For a mixed state $\rho_{A|B\textbf{C}}$ in a $2\otimes2\otimes2^{N-2}$ system, the following monogamy inequality for the $\alpha$th power of EoF holds:
\begin{equation}\label{eq:EoF}
E^{\alpha}_{A|B\textbf{C}}\geq E^{\alpha}_{AB}+E^{\alpha}_{A\textbf{C}},
\end{equation}
for $\alpha\geq\sqrt 2,$ and the following polygamy inequality holds:
\begin{equation}\label{eq:EoFv2}
E^{\alpha}_{A|B\textbf{C}}\leq E^{\alpha}_{AB}+E^{\alpha}_{A\textbf{C}},
\end{equation}
for $\alpha \leq 0.$

\emph{Proof:} Let's consider a tripartite pure state $|\phi_{AB\textbf{C}}\rangle$ in a $2\otimes2\otimes2^{N-2}$ system. Based on the Schmidt decomposition, the $2^{N-2}$-dimensional qubit $\textbf{C}$ can be viewed as an $effect$ four-dimensional qubit~\cite{Osborne06}. Therefore, we can consider the monogamy relationship in a $2\otimes2\otimes4$ system:
\begin{eqnarray}\label{eq:EoF_1}
E^{\alpha}_{A|B\textbf{C}}&=&\nonumber E^{\alpha}(\mathcal{C}^{2}_{A|B\textbf{C}})\\\nonumber
&\geq& E^{\alpha}(\mathcal{C}^{2}_{AB}+\mathcal{C}^{2}_{A\textbf{C}})\\\nonumber
&\geq& E^{\alpha}(\mathcal{C}^{2}_{AB})+E^{\alpha}(\mathcal{C}^{2}_{A\textbf{C}})\\
&=&E^{\alpha}(\rho_{AB})+E^{\alpha}(\rho_{A\textbf{C}}),
\end{eqnarray}
where the first inequality is due to $E(\mathcal{C}^2)$ is a monotonic increasing function and $\mathcal{C}^{2}_{A|BC}\geq \mathcal{C}^{2}_{AB}+\mathcal{C}^{2}_{A\textbf{C}}$ holds, the second inequality is due to the fact~\cite{Zhu14}: $E^{\alpha}(\mathcal{C}_1^2+\mathcal{C}_2^2)\geq E^{\alpha}(\mathcal{C}_1^2)+E^{\alpha}(\mathcal{C}_2^2)$ for all $\alpha\geq\sqrt 2$, the last equality is due to a mixed state $\rho_{AC}$ in a $2\otimes d$ system, $E(\rho_{A\textbf{C}})=E(\mathcal{C}^2(\rho_{A\textbf{C}}))$~\cite{Bai1402}. Thus, we complete our discussion on pure state.

Consider a mixed state $\rho_{AB\textbf{C}}$ in a $2\otimes2\otimes2^{N-2}$ system.  We use an optimal convex decomposition $\{p_i,|\phi^i_{AB\textbf{C}}\rangle \}$:
\begin{equation}
E(\rho_{A|B\textbf{C}})=\sum_ip_iE(|\phi^i_{AB\textbf{C}}\rangle),
\end{equation}
we can derive
\begin{eqnarray}
E(\rho_{A|B\textbf{C}})&=&\nonumber\sum_ip_iE(|\phi^i_{AB\textbf{C}}\rangle)\\\nonumber
&=&\sum_ip_iE[\mathcal{C}^2(|\phi^i_{AB\textbf{C}}\rangle)]\\\nonumber
&\geq& E[\sum_ip_i\mathcal{C}^2(|\phi^i_{AB\textbf{C}}\rangle)]\\\nonumber
&\geq& E[\mathcal{C}^2(\rho_{A|B\textbf{C}})]\\\nonumber
&\geq&\sqrt[\alpha]{E^{\alpha}(\mathcal{C}^{2}_{AB})+E^{\alpha}(\mathcal{C}^{2}_{A\textbf{C}})}\\
&=&E^{\alpha}(\rho_{AB})+E^{\alpha}(\rho_{A\textbf{C}}),
\end{eqnarray}
where the first equality is the definition of mixed state, we have used that $E(\mathcal{C}^2)$ is a convex function in the first inequality, the second inequality can be derived by Cauchy-Schwarz inequality: $(\sum_i x_i^2)^\frac{1}{2}(\sum_i y_i^2)^\frac{1}{2}\geq \sum_ix_iy_i,$ with $x_i=\sqrt{p_i}, y_i=\sqrt{p_i}C^2(|\phi^i_{AB\textbf{C}}\rangle).$ Thus proving the monogamy inequality. On the other hand, it is easy to check the polygamy inequality for $\alpha \leq 0.$ \qquad \qquad \qquad $\square$

Based on the discussion above, we show that for a mixed state $\rho_{A_1A_2\ldots A_N}$ in a $2\otimes2\otimes2^{N-2}$ system, a set of hierarchical monogamy inequalities holds for the $\alpha$th power of EoF in $k$-partite case with $k=\{3,4,\ldots,N\}$:
\begin{equation}
E^{\alpha}_{A_1|A_2\ldots{A_N}}\geq\sum_{i=2}^{k-1}E^{\alpha}_{A_1A_i}+E^{\alpha}_{A_1|A_k\ldots{A_N}},
\end{equation}
for $\alpha\geq\sqrt 2,$ which can be an improvement of Bai $et$ $al$'s work. And a set of hierarchical polygamy inequalities holds:
\begin{equation}
E^{\alpha}_{A_1|A_2\ldots{A_N}}\leq\sum_{i=2}^{k-1}E^{\alpha}_{A_1A_i}+E^{\alpha}_{A_1|A_k\ldots{A_N}},
\end{equation}
for $\alpha\leq0.$
When $k=N$, the general monogamy inequality hold:
\begin{equation}
E^{\alpha}_{A_1|A_2\ldots A_N}\geq E^{\alpha}_{A_1A_2}+...+E^{\alpha}_{A_1A_{N}},
\end{equation}
for $\alpha\geq\sqrt 2,$ the specific case have been revealed in Re.~\cite{Zhu14}. We also have the general polygamy inequality:
\begin{equation}
E^{\alpha}_{A_1|A_2\ldots A_N}\leq E^{\alpha}_{A_1A_2}+...+E^{\alpha}_{A_1A_{N}},
\end{equation}
for $\alpha\leq0.$

\section{Monogamy of $\alpha$th power Concurrence VS Monogamy of $\alpha$th power Negativity}\label{sec:Vs}
In this section, we will discuss the monogamy property of $\alpha$th power of concurrence and $\alpha$th power of negativity for $0< \alpha <2.$ Finally, we will discuss the monogamy property of $\alpha$th power of EoF for $0<\alpha<\sqrt 2.$

Based on the monogamy inequality of concurrence~\cite{Coffman00,Osborne06}, Re.~\cite{Zhu14} considered the general monogamy inequalities of $\alpha$th power concurrence in an $N$-qubit mixed state $\rho_{A_1A_2\cdots{A_N}}$, and claimed the following inequalities holds:
\begin{equation}
\mathcal{C}^{\alpha}_{A_1|A_2\cdots{A_N}}\geq \mathcal{C}^{\alpha}_{A_1A_2}+...+\mathcal{C}^{\alpha}_{A_1A_{N}}
\end{equation}
for $\alpha\geq 2$. While the polygamy inequalities holds:
\begin{equation}
\mathcal{C}^{\alpha}_{A_1|A_2\cdots{A_N}}< \mathcal{C}^{\alpha}_{A_1A_2}+...+\mathcal{C}^{\alpha}_{A_1A_{N}}.
\end{equation}
for all $\alpha\leq 0$. It's not clear for $0< \alpha <2$.

For convenience, we define the "residual tangle" of $\alpha$th power of concurrence as:
 \begin{equation}
\tau^\mathcal{C}(|\psi_{A_1A_2\ldots A_N}\rangle)=\mathcal{C}^{\alpha}_{A_1|A_2\ldots A_N}-\mathcal{C}^{\alpha}_{A_1A_2}-\cdots -\mathcal{C}^{\alpha}_{A_1A_N},
 \end{equation}
and define the "residual tangle" of $\alpha$th power of concurrence as:
\begin{equation}
\tau^\mathcal{N}(|\psi_{A_1A_2\ldots A_N}\rangle)=\mathcal{N}^{\alpha}_{A_1|A_2\ldots A_N}-\mathcal{N}^{\alpha}_{A_1A_2}-\cdots -\mathcal{N}^{\alpha}_{A_1A_N}.
 \end{equation}
Interestingly, We find that the $N$-qubit GHZ state
\begin{equation}
|GHZ\rangle=\frac{1}{\sqrt{2}}(|0\rangle^{\bigotimes N}+|1\rangle^{\bigotimes N}),
 \end{equation}
and $N$-qubit W state
\begin{equation}
|W\rangle=\frac{1}{\sqrt{N}}(|00\cdots 01\rangle+|00\cdots 10\rangle+\cdots +|10\cdots 00\rangle),
\end{equation}
can be used to distinguish the monogamous property of $\tau^\mathcal{C}(|\psi_{A_1A_2\ldots A_N}\rangle)$ for $0< \alpha <2$. In other words, $N$-qubit GHZ state is monogamous for the $\alpha$th power concurrence and $N$-qubit W state is polygamous for the $\alpha$th power concurrence, where $0< \alpha <2$. For $N$-qubit GHZ state
\begin{equation}
|GHZ\rangle=\frac{1}{\sqrt{2}}(|0\rangle^{\bigotimes N}+|1\rangle^{\bigotimes N}),
 \end{equation}
 the concurrence $\mathcal{C}_{A_1|A_2\ldots A_N}=1, \mathcal{C}_{A_1A_k}=0, k=\{2,3,\ldots ,N\}.$ Thus, the "residual tangle" $\tau^\mathcal{C}(|GHZ\rangle)=1>0$, $N$-qubit GHZ state is monogamous for the $\alpha$th power concurrence. For $N$-qubit W state
\begin{equation}
|W\rangle=\frac{1}{\sqrt{N}}(|00\cdots 01\rangle+|00\cdots 10\rangle+\cdots +|10\cdots 00\rangle),
\end{equation}
the concurrence $\mathcal{C}_{A_1|A_2\ldots A_N}=\frac{2}{N}\sqrt{N-1}, \mathcal{C}_{A_1A_k}=\frac{2}{N}, k=\{2,3,\ldots ,N\}.$ Thus, the "residual tangle" $\tau^\mathcal{C}(|W\rangle)=(\frac{2}{N})^{\alpha}[(N-1)^{\frac{\alpha}{2}}-(N-1)]<0$ for all $0< \alpha <2$. $N$-qubit W state is polygamous for the $\alpha$th power concurrence

For the "residual tangle" $\tau^\mathcal{N}(|\psi_{A_1A_2\ldots A_N}\rangle)$. The negativity of $N$-qubit GHZ state $\mathcal{N}_{A_1|A_2\ldots A_N}=1, \mathcal{N}_{A_1A_k}=0, k=\{2,3,\ldots ,N\}.$ Thus, $\tau^\mathcal{N}(|GHZ\rangle)=1>0$, it is coincide with $\tau^\mathcal{C}(|GHZ\rangle)$. The situation is different when we consider $\tau^\mathcal{C}(|\psi_{A_1A_2\ldots A_N}\rangle)$ for $N$-qubit W state. One obtain that $\mathcal{N}_{A_1A_2\ldots A_N}=\frac{2}{N}\sqrt{N-1}$ and $\mathcal{N}_{A_1A_k}=\frac{1}{N}\sqrt{2(N-2)^2+4-2(N-2)\sqrt{(N-2)^2+4}}, k=\{2,3,\ldots ,N\}.$ It is easy to check that $\tau^\mathcal{N}(|W\rangle)=\frac{1}{N^{\alpha}}[2^{\alpha}(N-1)^{\frac{\alpha}{2}}
-(N-1)[2(N-2)^2+4-2(N-2)\sqrt{(N-2)^2+4}]^{\alpha}]$. $\tau^\mathcal{N}(|W\rangle)$ can be positive and negative, as showed in Fig:1, we have plotted $\tau^\mathcal{N}(|W\rangle)$ as the function of $\alpha$ for $0<\alpha<2$, and consider $N=3$, $N=4$ and $N=5$ respectively. We find $\tau^\mathcal{N}(|W\rangle)$ is not always negative, which is different than the case of $\tau^\mathcal{C}(|W\rangle)$.

\begin{figure}[htbp]
  \centering
 \includegraphics[trim = 0MM 0MM 0MM 0MM, clip=true, width=8CM,height=8CM]{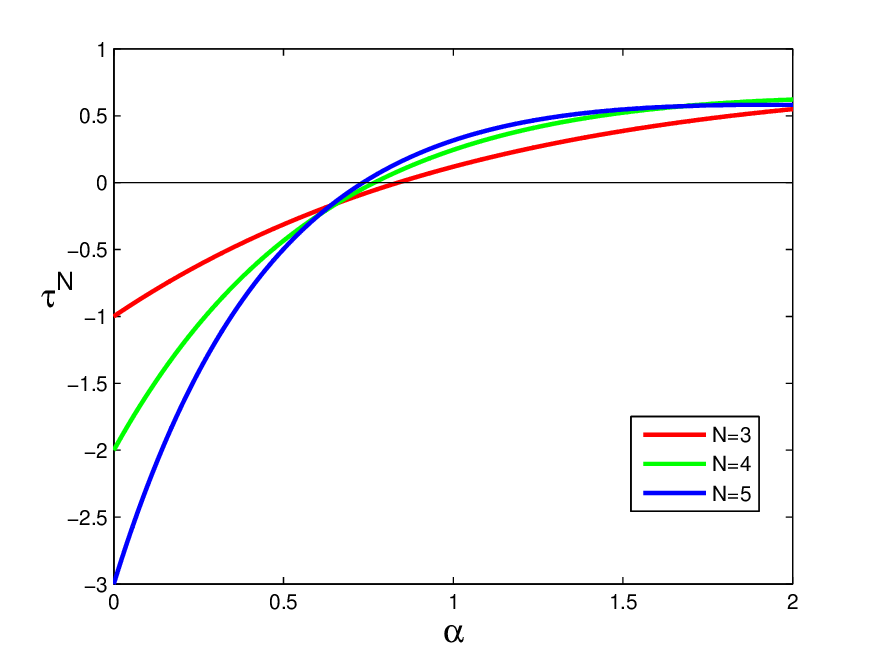}
  \caption{(Color online) $\tau^N(|W\rangle)$ as the function of $\alpha$ for $0<\alpha<2$, the red line, green line and blue line denote $N=3$, $N=4$ and $N=5$ respectively.}\label{Fig:1}
\end{figure}

Finally, we will discuss the monogamy property of $\alpha$th power of EoF for $0<\alpha<\sqrt 2$. We define the "residual tangle" of $\alpha$th power of EoF as:
\begin{equation}
\tau^E(|\psi_{A_1A_2\ldots A_N}\rangle)=E^{\alpha}_{A_1|A_2\ldots A_N}-E^{\alpha}_{A_1A_2}-\cdots -E^{\alpha}_{A_1A_N}.
\end{equation}

For $N$-qubit GHZ state, the EoF of $N$-qubit GHZ state $E_{A_1|A_2\ldots A_N}=1,$ $E_{A_1A_k}=0,$ $k=\{2,3,\ldots,N\}.$ Thus, the "residual tangle" $\tau^E(|GHZ\rangle)=1>0$ for $0<\alpha<\sqrt 2.$  For $N$-qubit W state, the EoF of $N$-qubit W state $E_{A_1|A_2\ldots A_N}=h(\frac{1+\sqrt{1-4\frac{N-1}{N^2}}}{2})=h(1-\frac{1}{N}),$ $E_{A_1A_k}=h(\frac{1+\sqrt{1-\frac{4}{N^2}}}{2})=h(\frac{N+\sqrt{N^2-4}}{2N})$ for $k=\{2,3,\ldots,N\},$ where $h(x)$ denotes the binary function. Thus, the "residual tangle" $\tau^E(|W\rangle)=h^{\alpha}(1-\frac{1}{N})-(N-1)h^{\alpha}(\frac{N+\sqrt{N^2-4}}{2N})$ where $0<\alpha<\sqrt 2$ and $N\geq3.$ We have proved in the appendix $\tau^E(|W\rangle)<0$ for $0<\alpha\leq\frac{1}{2}$. On the other hand, $\tau^E(|W\rangle)$ can be positive and negative for $\frac{1}{2}<\alpha<\sqrt 2$. As showed in Fig:2, we have plotted $\tau^\mathcal{E}(|W\rangle)$ as the function of $\alpha$ for $0<\alpha<\sqrt2$, and consider $N=3$, $N=4$ and $N=5$ respectively.

\begin{figure}[htbp]
  \centering
 \includegraphics[trim = 0MM 0MM 0MM 0MM, clip=true, width=8CM,height=8CM]{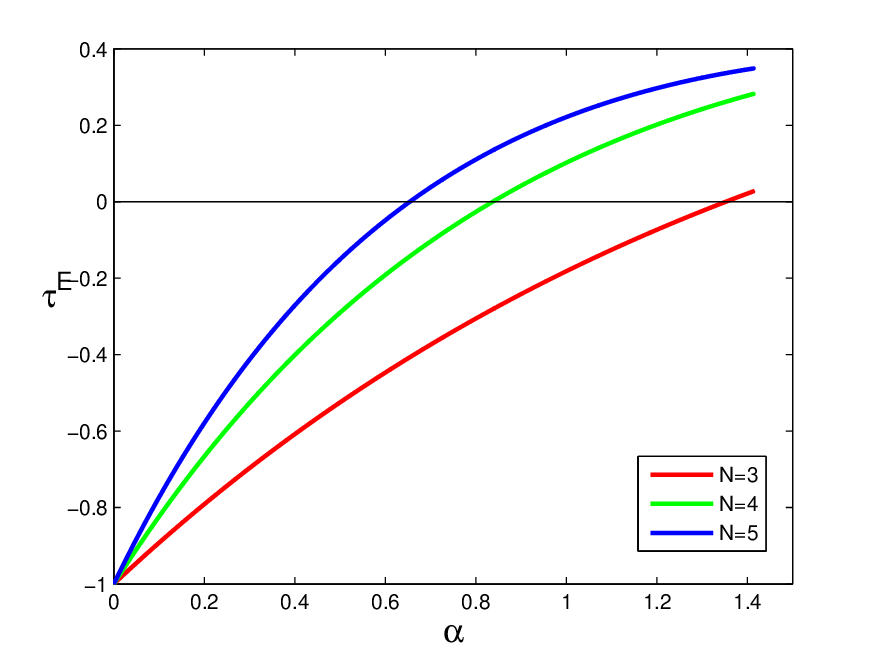}
  \caption{(Color online) $\tau^E(|W\rangle)$ as the function of $\alpha$ for $0<\alpha<\sqrt 2$, the red line, green line and blue line denote $N=3$, $N=4$ and $N=5$ respectively.}\label{Fig:2}
\end{figure}

\section{conclusion}\label{sec:conclusion}
In this paper, We studied the monogamy property of $\alpha$th power of entanglement measure in bipartite states. In particular, we investigated the monogamy properties of negativity and CREN in detail. We showed that the $\alpha$th power of negativity, CREN are monogamous for $\alpha\geq2$ and polygamous for $\alpha\leq0$. We improved the hierarchical monogamy inequality for the $\alpha$th power of EoF, and show that the $\alpha$th power of EoF is hierarchical monogamous for $\alpha\geq\sqrt2$. Finally, we discussed the monogamy property of $\alpha$th power of concurrence and the $\alpha$th power of EoF. We found that the $N$-qubit GHZ state and $N$-qubit W state can be used to distinguish both the $\alpha$th power of the concurrence for $0<\alpha<2$ and the $\alpha$th power of the EoF for $0<\alpha<frac{1}{2}$ in qubit system. We compared concurrence with negativity in terms of monogamy property and showed the difference between them.

\section{acknowledgments}
We thank Zheng-Jun Xi and Chen-Ming Bai for their helpful discussions. Y. Li was supported by NSFC (Grants No.11271237 and No.61228305) and the Higher School Doctoral Subject Foundation of Ministry of Education of China (Grant No.20130202110001). Y. Luo was supported by NSFC (Grant No.61303009).

\section{appendix}
For a binary entropy function $h(x)$ for $0\leq\alpha\leq1$. We have following lower bounding and upper bounding for approximation~\cite{Calabro09}:
\begin{equation}
1-4(x-\frac{1}{2})^2\leq h(x)\leq1-\frac{(1-2x)^2}{2\ln 2},
\end{equation}
for $0\leq x\leq1$.

The "residual tangle" is:
\begin{eqnarray}
\tau^E(|W\rangle)&=&\nonumber h^{\alpha}(1-\frac{1}{N})-(N-1)h^{\alpha}(\frac{N+\sqrt{N^2-4}}{2N})\\\nonumber 
&\leq& [1-\frac{(1-\frac{2}{N})^2}{2\ln 2}]^{\alpha}-(N-1)[1-4(\frac{\sqrt{N^2-4}}{2N})^2]^{\alpha}\\\nonumber
&<& 1-(N-1)(\frac{2}{N})^{2\alpha},\\
\end{eqnarray}
where $0<\alpha<\sqrt 2$ and $N\geq3$. To prove $\tau^E(|W\rangle)<0$ for $0<\alpha\leq\frac{1}{2}$ and $N\geq3$, we define:
\begin{equation}
f(x)=(x-1)(\frac{2}{x})^{2\alpha},
\end{equation}
where $x\geq3$ is a real number and $0<\alpha\leq\frac{1}{2}$. The derived function of $f(x)$ is:
\begin{equation}
f'(x)=(\frac{2}{x})^{2\alpha}[\frac{(1-2\alpha)x+2\alpha}{x}].
\end{equation}

It is easy to check that for $0<\alpha\leq\frac{1}{2}$, the derived function $f'(x)\geq0$. Thus $f(x)$ is monotonic increasing, which derived $\tau^E(|W\rangle)$ is monotonic decreasing for $N$. The maximum of $\tau^E(|W\rangle)$ is $\max\tau^E(|W\rangle)=1-f(3)=1-2(\frac{2}{3})^{2\alpha}<0$.

\end{document}